\documentclass[12pt]{iopart}
\usepackage{graphicx,color}
\usepackage{subfigure}
\usepackage{verbatim}
\usepackage{ifpdf}
\usepackage{ifthen} 
\usepackage{url}

\ifpdf
\usepackage[pdftex]{pdflscape}
\usepackage[pdftex,colorlinks=true]{hyperref}

\else
\usepackage{lscape}
\usepackage[dvips,colorlinks=false]{hyperref}

\fi
\usepackage{breakurl}

\hypersetup{
pdftitle={A Comparison of Finite Element and Atomistic Modelling of Fracture},
pdfauthor={Valerie R. Coffman, James P. Sethna, Gerd Heber, Mu Liu, Anthony
Ingraffea, Nicholas P. Bailey, Erin Iesulauro Barker},
pdfkeywords={grain boundaries, molecular dynamics, fracture mechanics, finite
  elements, solid state physics,
  computational physics},
bookmarksnumbered,
pdfstartview={FitH},
pdfpagemode=UseOutlines,
breaklinks=true
}

\newcommand{\mycaption}[2][]{
  \ifthenelse{\equal{}{#1}} {\caption{#2}} {\caption[#1]{ {\bf#1} #2}}}
\newcommand{\inThesis}[1]{}
\newcommand{\inPaper}[1]{#1}
\newcommand{\onlinecite}[1]{\cite{#1}}

\begin{document}

\title{A comparison of finite element and atomistic modelling of fracture}
\date{\today}

\author{V R Coffman$^1$\footnote{Present address: Information Technology
Laboratory, National Institute of Standards and Technology, Gaithersburg, MD,
20899 }, J P Sethna$^1$, G Heber$^2$, M Liu$^2$, A Ingraffea$^2$, N P Bailey$^3$,
E I Barker$^4$ }
\address{$^1$ Laboratory of Atomic and Solid State Physics (LASSP), Clark Hall,
Cornell University, Ithaca, NY 14853-2501, USA}
\address{$^2$ Cornell Fracture Group, Rhodes Hall,
Cornell University, Ithaca, NY 14853-2501, USA}
\address{$^3$ Department of Mathematics and Physics (IMFUFA), DNRF Center
``Glass and Time'', Roskilde University, P.O. Box 260, DK-4000 Roskilde, Denmark}
\address{$^4$Los Alamos National Laboratory, Los Alamos, New Mexico 87545, USA}
\ead{valerie.coffman@nist.gov}

\date{\today}

\begin{abstract}

Are the cohesive laws of interfaces sufficient for modelling fracture in
polycrystals using the cohesive zone model?  We examine this question by
comparing a fully atomistic simulation of a silicon polycrystal to a finite
element simulation with a similar overall geometry.  The cohesive laws used in
the finite element simulation are measured atomistically.  We describe in detail
how to convert the output of atomistic grain boundary fracture simulations into
the piecewise linear form needed by a cohesive zone model.  We discuss the effects of grain
boundary microparameters (the choice of section of the interface, the
translations of the grains relative to one another, and the cutting plane of
each lattice orientation) on the cohesive laws and polycrystal fracture.  We
find that the atomistic simulations fracture at lower levels of external stress,
indicating that the initiation of fracture in the atomistic simulations is
likely dominated by irregular atomic structures at external faces, internal
edges, corners, and junctions of grains. Thus, cohesive properties of interfaces
alone are likely not sufficient for modelling the fracture of polycrystals using
continuum methods.

\end{abstract}


\submitto{\MSMSE}

\maketitle

\section{Introduction}

The cohesive zone model~\cite{needleman-interfaceCZM} (CZM), a finite element
based method for simulating fracture, is often applied to polycrystals and
multiphase materials which fracture at the interfaces of grains or material
phases.  The debonding of such interfaces is described by cohesive laws which
give the displacement across the interfaces as a function of stress. An example
of a cohesive law is shown in Figure~\ref{fig:piecewiseexample}.  It has been
shown that the shape and scale of the cohesive law has a large effect on the
outcome of the finite element simulation~\cite{needleman-interfaceCZM,
falk-critical}.  However, previous CZM simulations of polycrystals have used
cohesive laws that are guessed, chosen for numerical convergence, and do not
take into account the effect of varying grain boundary geometries within the
material.  The same cohesive law is often used throughout the material despite
the fact that in a real material, the geometries of the grain boundaries/phase
interfaces must vary~\cite{iesulauro1, iesulauro2, warner-sliding-plasticity}.

For input into CZM simulations of polycrystals, it would be useful to find a
formula for the cohesive laws of grain boundaries as a function of their
geometry.  Numerous studies have shown that there are large jumps in the peak
stress for special grain boundaries~\cite{chen-GB, old-grain-boundary-sims,
pumphrey, sansoz-molinari-structure, shenderova-diamond, wolf-structure,
wolf-merkle, wolf-jaszczak}.  A recent systematic study of 2D grain boundaries
has shown that perturbing special, commensurate grain boundaries adds nucleation
sites for fracture, lowering the fracture strength of the
boundary~\cite{grain-boundary-geo}.  This leads to a hierarchical structure to
the fracture strength as a function of geometry, with singularities at all
commensurate grain boundaries.

\begin{figure}[thb]
\subfigure[]{\begin{minipage}[c]{0.42\linewidth}
\includegraphics[width=6cm]{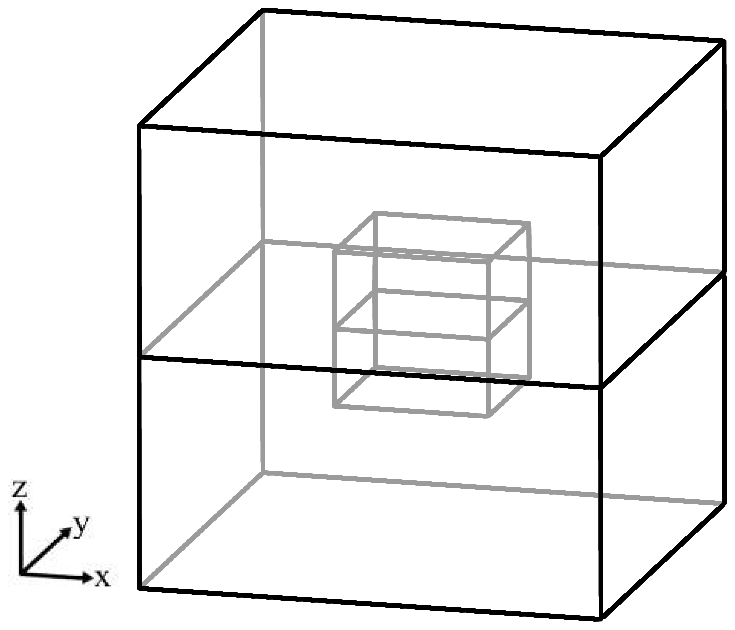}
\end{minipage}}
\subfigure[]{\begin{minipage}[c]{0.58\linewidth}
\includegraphics[height=2.5cm]{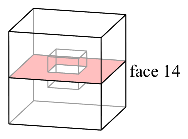}
\includegraphics[height=2.5cm]{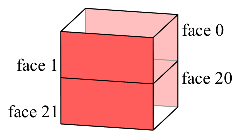}

\includegraphics[height=2.5cm]{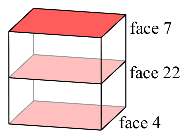}
\includegraphics[height=2.5cm]{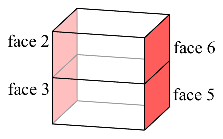}
\label{fig:CubeInCubeFaces}
\end{minipage}}
\mycaption[Schematic Diagram of the Cube-In-Cube Model.]  {Figure (a) shows a
  schematic diagram of the cube-in-cube model. The inner cube is centered within
  the outer cube and has a length equal to 1/3 that of the outer cube.  Figure (b)
  shows the numbering of the internal faces of the model.  The upper-left figure
  shows the entire cube-in-cube model, while the rest show only the inner cube.  In
  our simulation, we load the upper face of the model in the $z$-direction.
  Under such loading, faces 4, 7, 14, and 22 are subject to pure normal
  traction.  Faces 0, 1, 2, 3, 5, 6, 20, and 21, are subject to pure shear
  traction. (The numbering of the internal faces is not contiguous because the
  finite element simulation also numbers the ten external faces.)  The inner cube is a
  single crystal, but in order to allow for intragranular fracture through this
  crystal, we add an internal face through the center.  The constitutive
  relation for this interface is that of a perfect crystal.  Notice that pairs
  0\&1, 2\&6, 3\&5, and 20\&21 are boundaries that macroscopically have
  identical cohesive laws since they are related by an inversion, i.e.~they
constitute symmetric pairs of interfaces for
  which the grains have been swapped.}
\label{fig:CubeInCube} 
\end{figure}

Since finding a functional form for 3D grain boundary cohesive laws is daunting,
it is helpful to calculate the cohesive laws atomistically, on the fly, for each
geometry in a given polycrystal.  (It is less feasible to measure cohesive laws
experimentally because it is difficult to isolate and measure the displacements
on either side of the grain boundary.)  But are the cohesive laws of the
interfaces enough? In this \inThesis{chapter}\inPaper{paper}, we will compare a
finite element, cohesive zone model that uses an atomistically generated
cohesive law for each interface to a fully atomistic simulation of the same
geometry.  We will compare the stress fields of each model and the overall
pattern of fracture.  The model we will investigate is that of a cube embedded
in a boundary that bisects a larger cube (Figure~\ref{fig:CubeInCube}) with a
normal load imposed on the top face.  The model has three regions, the two
halves of the outer cube, and the inner cube, each with a different lattice
orientation.  The orientations are chosen at random and shown in
Table~\ref{table:latticeorientations}.  We
calculate a cohesive law atomistically for each interface of the model. To
allow for intragranular fracture through the inner cube, we have added an
interface through the center.  For this interface, we measure the cohesive law
of a perfect crystal.

\begin{table*}[thb]
\caption[Lattice Orientations]{ The lattice orientations of the three regions of
the cube-in-cube model are given below as Euler angles, expressed in degrees.}
\label{table:latticeorientations}
\begin{indented}
\item[]
\begin{tabular}{ @{} l p{5.2cm} r@{.}l r@{.}l r@{.}l }
\br
&& \multicolumn{2}{l}{Center Cube} & 
\multicolumn{2}{l}{Upper Half} & 
\multicolumn{2}{l}{Lower Half} \\
\mr
$\theta_z$ & 
First rotation about the $z$-axis (positive $x$-axis to $y$)
& 27&80 & -79&28 & 34&09 \\
$\theta_x$ & 
Second rotation about the intermediate local $x$-axis (positive $y$-axis to $z$)
& 27&65 & 66&52 & 27&40 \\
$\theta_y$ &
Third rotation about intermediate local $y$-axis (positive $z$-axis to $x$)
& 89&49 & 23&64 & -73&79 \\
\br
\end{tabular}
\end{indented}
\end{table*}

\section{The Cohesive Zone Model}
\label{sec:CZM}
\begin{figure}[thb]
\begin{center}
\subfigure[]{
\includegraphics[width=4cm]{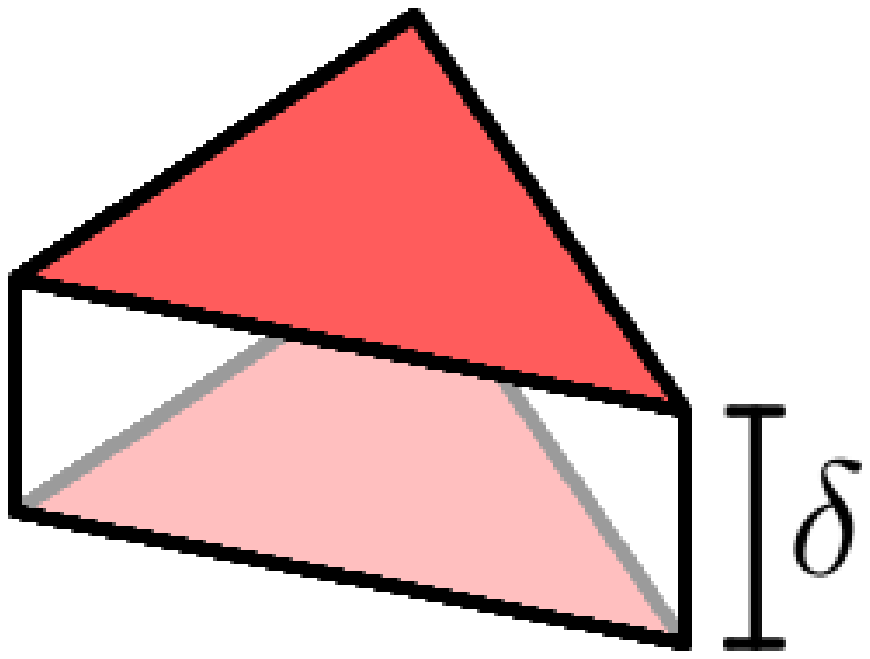}
\label{fig:interfaceelement}}
\subfigure[]{
\includegraphics[width=6cm]{cube_in_cube_figures/PiecewiseExample}
\label{fig:piecewiseexample}}
\end{center}
\mycaption[Interface Elements and the Piecewise Linear Cohesive Law.]{ Figure
(a) shows a schematic diagram of a triangular interface element.  The displacement across
the interface $\delta$, is initially zero.  Each of the two triangles forms a
face of one of the tetrahedral elements in the material on either side of
the interface.  Figure (b) shows the form of the constitutive relation for the
interface elements.  The slope of the first linear segment is the initial
stiffness, $k_0$.  When the traction across the interface reaches the peak
traction, $\tau_p$, the interface element begins to soften.  When the normalized
displacement, defined by $\lambda=\delta/\delta_c$ reaches a value of 1, the
interface has fully debonded.}
\end{figure}

The cohesive zone constitutive model is implemented in a finite element model
with zero volume interface elements placed between the regular finite elements
at interfaces. An example of an interface element is shown in
Figure~\ref{fig:interfaceelement}. These interface elements simulate fracture by
debonding according to a cohesive law, the relation between the traction and
displacement across the interface.  The form of cohesive law used here is the
piecewise linear form developed by Tvergaard and
Hutchinson~\cite{tvergaard-hutchinson} also described by Gullerud \textit{et
al}.~\cite{warp3dmanual}.  An example is shown in
Figure~\ref{fig:piecewiseexample}.  The piecewise linear form of the cohesive
law is determined by the initial stiffness $k_0$, the peak traction $\tau_p$,
and the critical displacement, $\delta_c$ at which the surface is considered
fully debonded and traction-free.

Camacho and Ortiz~\cite{camacho-ortiz} describe mixed loading by assigning
different weights to the tangential and normal components of displacement,
described by a factor $\beta$.  We also assume that the resistance relative to
tangential displacements is considered to be independent of direction.  This
leads to an effective displacement of
\begin{equation}
\delta = \sqrt{ \delta_n^2 + \beta  \delta_t^2 }
\end{equation}
where $\delta_n$ is the normal displacement and $\delta_t$ is the tangential
displacement.  The effective traction is
\begin{equation}
\tau = \sqrt{ \tau_n^2 + \beta^{-2} \tau_t^2 }
\end{equation}
where $\tau_n$ is the normal component of traction and $\tau_t$ is the
tangential component of traction.

\section{Atomistically Determined Material Properties Used by CZM}

The parameters needed by the CZM simulation that are determined by atomistics
are the elastic constants associated with the atomic potential, the orientation
of the lattice in each grain, and the cohesive law of each interface.  We are
modelling silicon using the Stillinger-Weber (SW)
potential~\cite{stillinger-weber} but with an extra parameter, $\alpha$,
multiplying the three body term such that $\alpha=1$ corresponds to standard SW.
We use two values for this parameter, the standard value of 1.0 (matching to
other properties of real silicon) and a value of 2.0 which makes the material
more brittle.  We calculate separate material properties (elastic constants and
cohesive laws) for each version of SW.

\subsection{Determining the Elastic Constants}
\label{sec:measuring_elastic_constants}
In order to make a direct comparison between the atomistic and finite element (FE) simulations,
we must determine the elastic constants of each version of SW for
input into the FE simulation.  The elastic constants are measured by
initializing a cube of atoms in a diamond lattice, incrementing a strain in one
direction, relaxing the atoms at zero temperature, and measuring the stress tensor
at each increment.

For an interatomic potential with 3-body terms of the form
\begin{equation}
E_i = \sum_{j<k} f(\vec{r_{ij}}, \vec{r_{ik}}),
\end{equation}
we can find the $\alpha \beta$ component of stress at atom $i$ by utilizing the
relation~\cite{HirthAndLothe}
\begin{equation}
(\sigma_i)_{\alpha,\beta}
   = \frac{1}{V} \frac{\partial E_i}{ \partial \epsilon_{\alpha, \beta}}
\end{equation}
where $V$ is the volume per atom. This leads
to
\begin{equation}
(\sigma_i)_{\alpha,\beta}
   = \frac{1}{V} \sum_{j<k} \frac{\partial E_i}{ \partial \vec{r_{ij}} } \cdot \frac
   {\partial \vec{r_{ij}} }{ \partial \epsilon_{\alpha, \beta} } + \frac
   {\partial E_i }{ \partial \vec{r_{ik}}} \cdot \frac {\partial \vec{r_{ik}} }{
   \partial \epsilon_{\alpha, \beta}}.
\end{equation}
Because 
\begin{equation}
\partial (r_{ij})_\gamma / \partial \epsilon_{\alpha, \beta}  = (r_{ij})_\beta
\delta_{\alpha,\gamma},
\end{equation}
the atomic stress is
\begin{equation}
 (\sigma_i)_{\alpha,\beta}  = \frac{1}{V} \sum_{j<k} \frac{ \partial E_i }{ \partial (r_{ij})_\alpha }
   (r_{ij})_\beta + \frac {\partial E_i }{ \partial (r_{ik})_\alpha}
   (r_{ik})_\beta.
\label{eqn:atomicstress}
\end{equation}
We use a value of $V$ equal to the volume per atom in the ground state (perfect
lattice).   This has the shortcoming that for atoms near dislocations or grain boundaries,
the actual volume per atom will be quite different.

$C_{11}$, $C_{12}$, and $C_{44}$ are determined by
$\sigma_{xx}/\epsilon_{xx}$, $\sigma_{xx}/\epsilon_{yy}$, and
$\sigma_{xy}/\epsilon_{xy}$, respectively.  The results are given in Table
\ref{table:elasticConstantsOfSilicon}.  The differences between the elastic
constants calculated for SW silicon and those found by experiement
are due to the fact that the inter-atomic potential is an approximate representation
of real silicon.

\begin{table*}[thb]
\caption{Elastic constants of silicon, determined atomistically for two versions
of SW silicon.}
\label{table:elasticConstantsOfSilicon}
\lineup
\begin{indented}
\item[]
\begin{tabular}{@{}l r@{.}l  r@{.}l  l}
\br
 & \multicolumn{2}{l}{Original SW (GPa)}& 
 \multicolumn{2}{l}{Brittle SW (GPa)} &
 \multicolumn{1}{l}{Experiment~\cite{deLaunay} (GPa)}\\
\mr
$C_{11}$ & 69&74 & 92&78 & 166 \\
$C_{12}$ & 35&20 & 23&69 & \064 \\
$C_{44}$ & 52&00 & 83&37 & \080 \\
\br
\end{tabular}
\end{indented}
\end{table*}

\subsection{Calculating the Cohesive Laws}
\label{sec:MeasuringCohesiveLawForCubeInCube}

\begin{figure}[thb]
\begin{center}
\includegraphics[width=10cm]{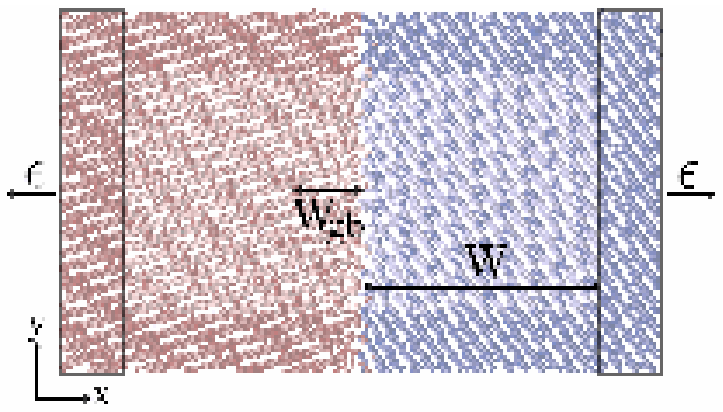} \mycaption[An
Example of a Grain Boundary Simulation.]{The darker atoms represent constrained
layer of atoms used to enforce rollered boundary conditions.  The normal strain
is imposed by displacing the constrained atoms on the endcaps which are indicated by the
rectangles.  $W$ is the width of the unconstrained atoms in each grain in
the direction perpendicular to the interface.  $W_{gb}$ represents the width of
the strain field on either side of the interface, and is chosen such 
that~\eref{eqn:subtract_elastic} does not give a negative value. }
\label{fig:3DGBExample}
\end{center}
\end{figure}

\begin{figure}[thb]
\begin{center}
\includegraphics[width=10cm]{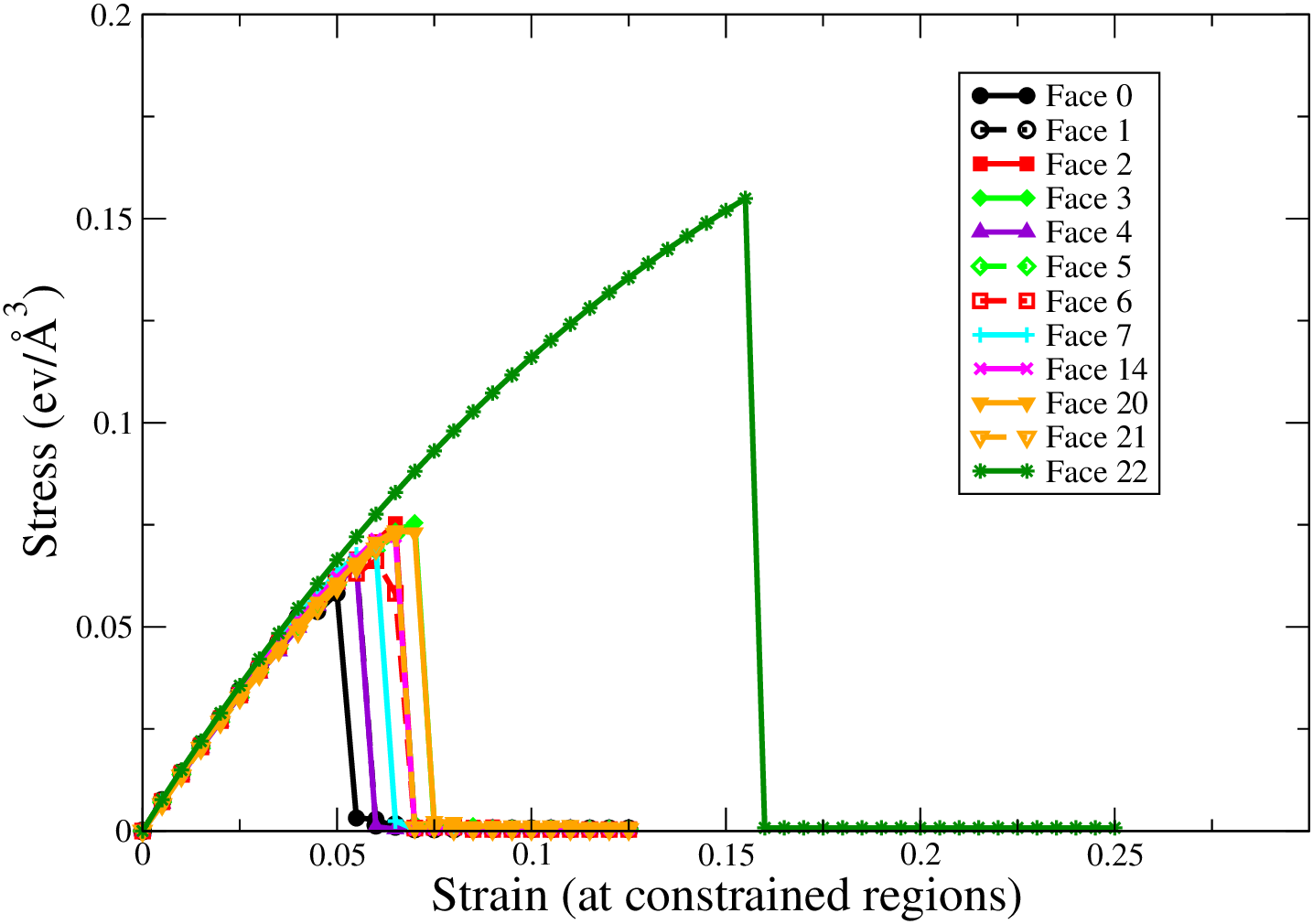}
\mycaption[Strain versus Stress: Brittle SW Silicon.]
{Strain versus stress for the twelve different interfaces needed for the
continuum FE cube-in-cube simulation. Each grain is 30\AA\ on a side.  In order
to calculate the cohesive law for only the grain boundary, we subtract off the elastic
response of the bulk (Figure~\ref{fig:CohesiveBrittleCorrected}).  Face 0
has no grain boundary, representing intragranular fracture through the center
cube.  Notice that the cohesive laws of the vertical interfaces are invariant under
inversion, so some pairs of faces would have identical cohesive laws if measured
in an infinite-sized system (or one with periodic boundary conditions and
micro-parameters that are completely optimized). Thus the differences between faces 0\&1,
2\&6, 3\&5, and 20\&21, both here and in
Figures~\ref{fig:CohesiveBrittleCorrected},\ref{fig:BilinearCohesiveBrittle},
and \ref{fig:BilinearCohesiveOrdinary} reflect the discreteness effects of the
choice of lattice origin and positions of the edges of the simulation.}
\label{fig:CohesiveBrittleRaw} 
\end{center}
\end{figure}

The method for computing the cohesive law of a grain boundary with an atomistic
simulation is described in\inThesis{
section~\ref{sec:measuring_cohesive_laws}}\inPaper{~\onlinecite{grain-boundary-geo}}.
Here we are measuring fully 3D grain boundary geometries.  An example of a 3D
grain boundary simulation is shown in Figure~\ref{fig:3DGBExample}.  In order to
simulate grain boundaries of any geometry (not just geometries restricted by
commensurability) we use constrained layers of atoms on the surfaces instead of
periodic boundary conditions.  Because SW contains three body terms, a layer
thickness equal to twice the cutoff distance of the potential is needed to
ensure that the free atoms are not subject to surface effects.  There is a
constrained zone for each face (atoms that are within a constrained zone width
of a single exterior face), edge (atoms that are within a constrained zone width
of exactly two exterior faces), and corner (atoms that are within a constrained
zone width of three exterior faces).  The atoms on the faces are constrained to
not move perpendicular to the face, the atoms on the edge are constrained to
move only parallel to the edge, and the atoms in the corners are totally fixed
in position.  These constraints simulate ``rollered'' boundary conditions.  Each
grain is 30 $\mbox{\AA}$ wide and normal strain increments are 0.005. At each
strain increment, we displace ``endcaps'', the constrained atoms adjacent to the
external faces that are parallel to the $yz$-plane (indicated by the gray boxes
in Figure~\ref{fig:3DGBExample}).  We relax the atoms and measure the $xx$
component of stress on the endcaps.  An example of the stress-strain curve that
results from such a simulation is shown in Figure~\ref{fig:CohesiveBrittleRaw}.

The CZM uses a traction-displacement law that describes the debonding at the
interface in question~\cite{xu-needleman,camacho-ortiz,tvergaard-hutchinson} as
discussed in section~\ref{sec:CZM}.  The piecewise linear form is determined by
the initial stiffness $k_0$, the peak traction $\tau_p$, and the final
displacement $\delta_c$.  We will need to extract these parameters from the
output of our grain boundary simulation (Figure~\ref{fig:CohesiveBrittleRaw}).

\begin{figure}[thb]
\begin{center}
\includegraphics[width=10cm]{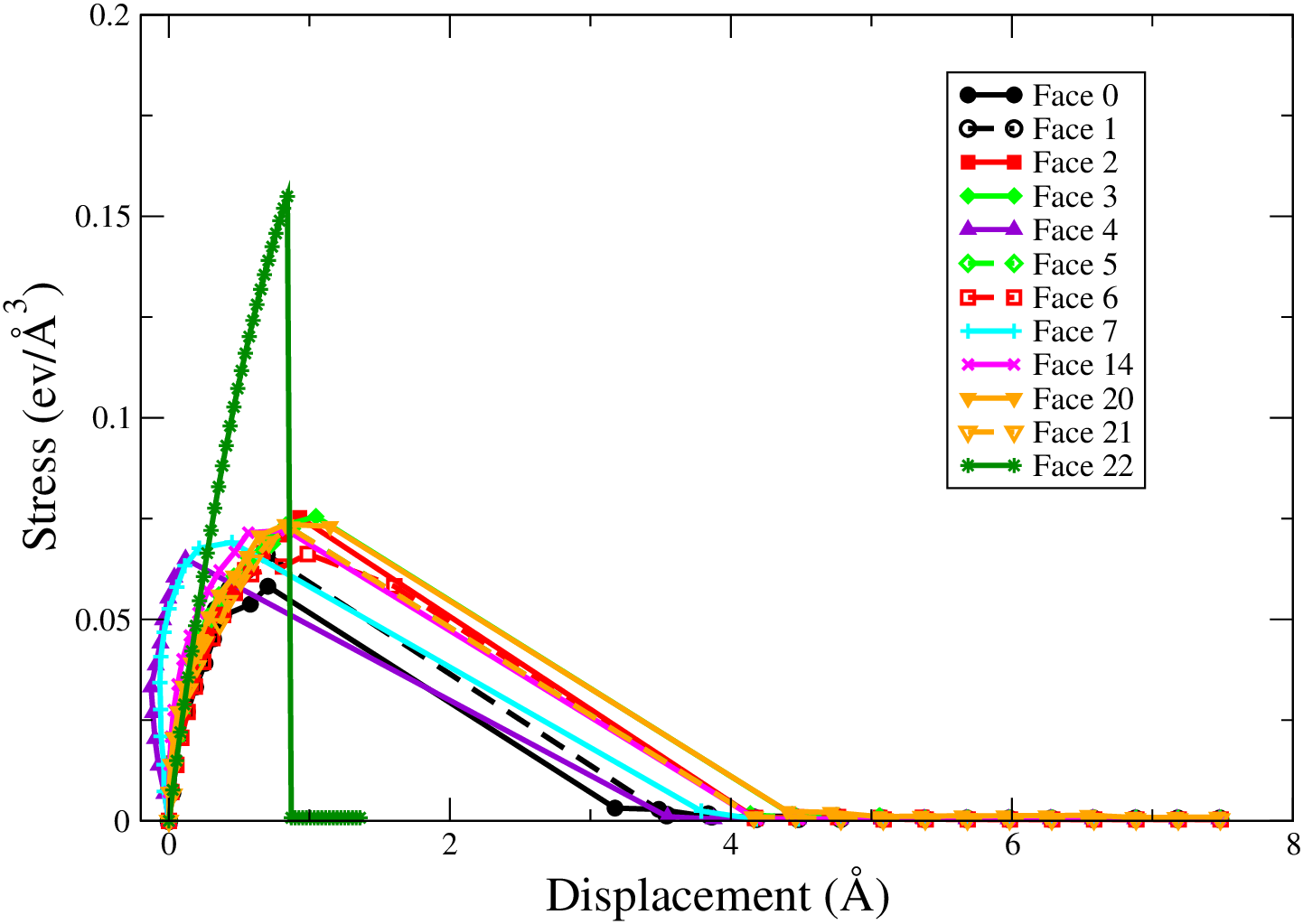}
\mycaption[Cohesive Law: Brittle SW Silicon.]
{Cohesive law, displacement versus stress, for the brittle potential and twelve
  interfaces of Figure~\ref{fig:CohesiveBrittleRaw}. The transformation from
  strain to effective displacement at the interface is as described in
  section~\ref{sec:MeasuringCohesiveLawForCubeInCube}. The effective thickness
  of the interface is 9\AA\ on each side. (Note that this is comparable to the
  entire size of the smaller atomistic cube-in-cube simulations.)}
\label{fig:CohesiveBrittleCorrected}
\end{center}
\end{figure}

\begin{figure}[thb]
\begin{center}
\subfigure{
\includegraphics[width=7cm]{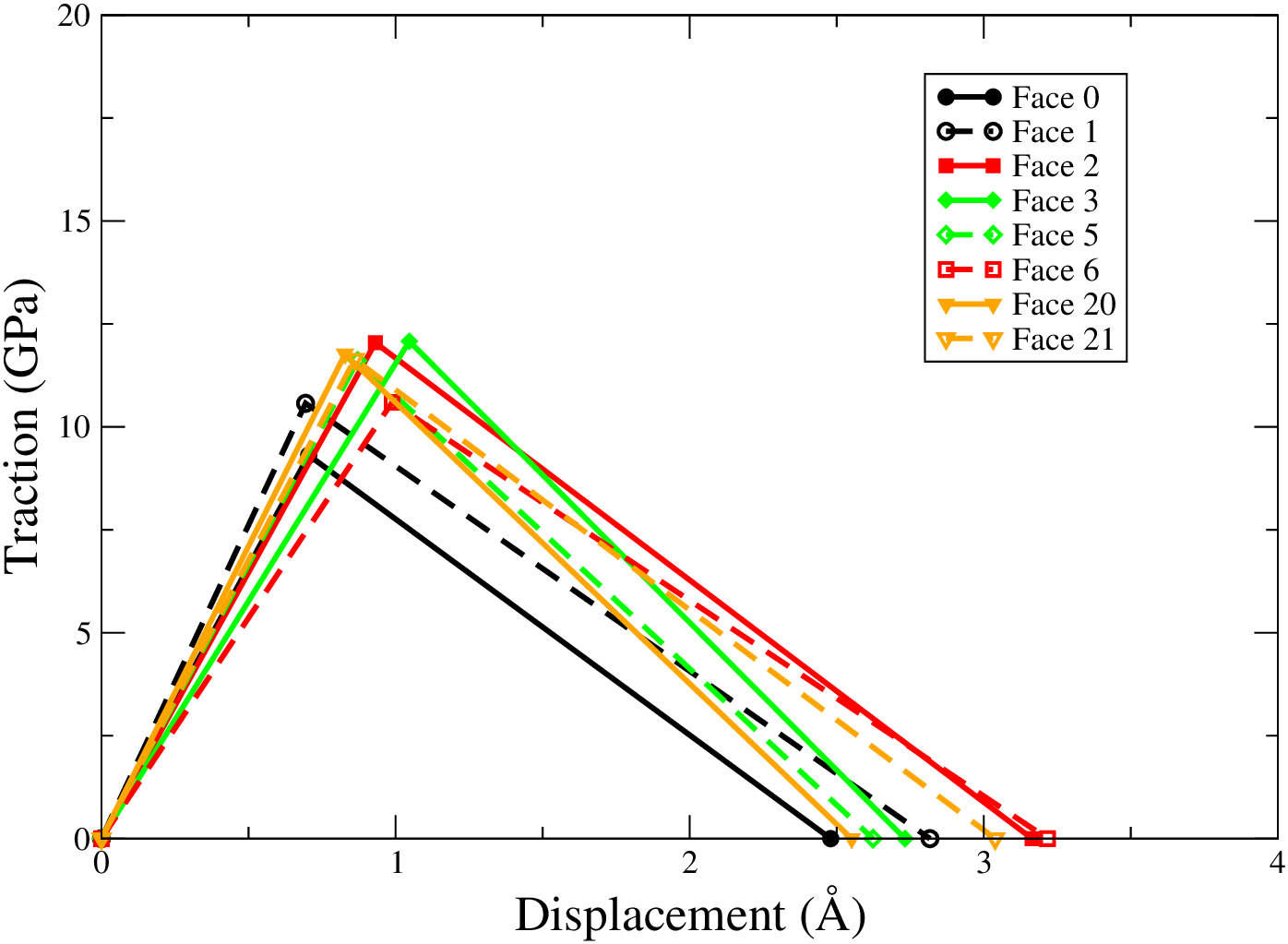}}
\subfigure{
\includegraphics[width=7cm]{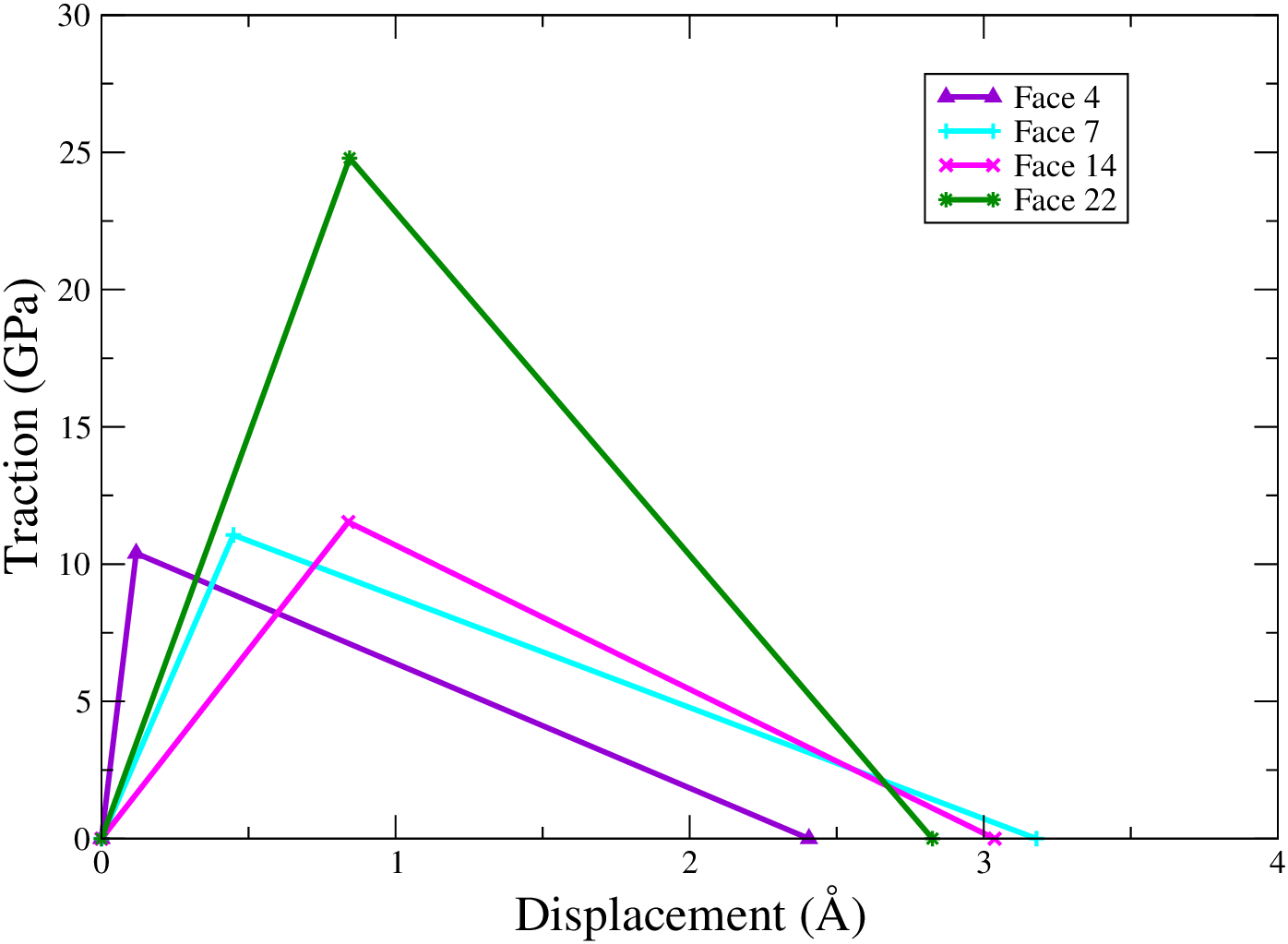}}
\mycaption[Piecewise Linear Cohesive laws: Brittle SW Silicon.]
{Simplified piecewise linear cohesive law used in the FE simulations. The peak
stress and its corresponding displacement were taken from Figure
\ref{fig:CohesiveBrittleCorrected}, and the critical displacement where the
force vanishes is chosen to make the area under the curve equal the Griffith
energy. In the figure on the left, solid and dashed line pairs of the same shade
indicate pairs of interfaces with the same macroparameters (grain orientations)
but different microparameters.}
\label{fig:BilinearCohesiveBrittle}
\end{center}
\end{figure}

\begin{figure}[thb]
\begin{center}
\subfigure{
\includegraphics[width=7cm]{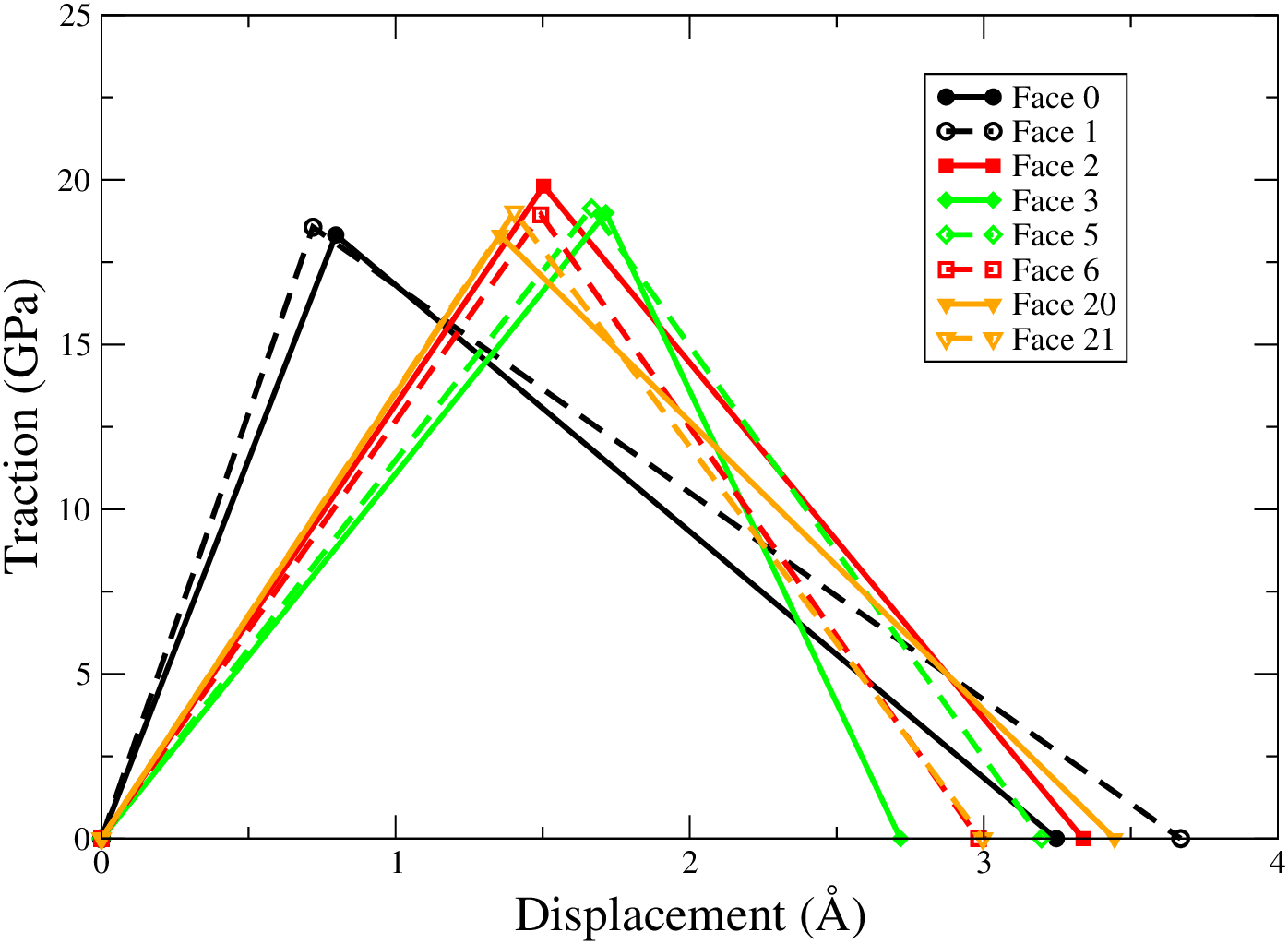}}
\subfigure{
\includegraphics[width=7cm]{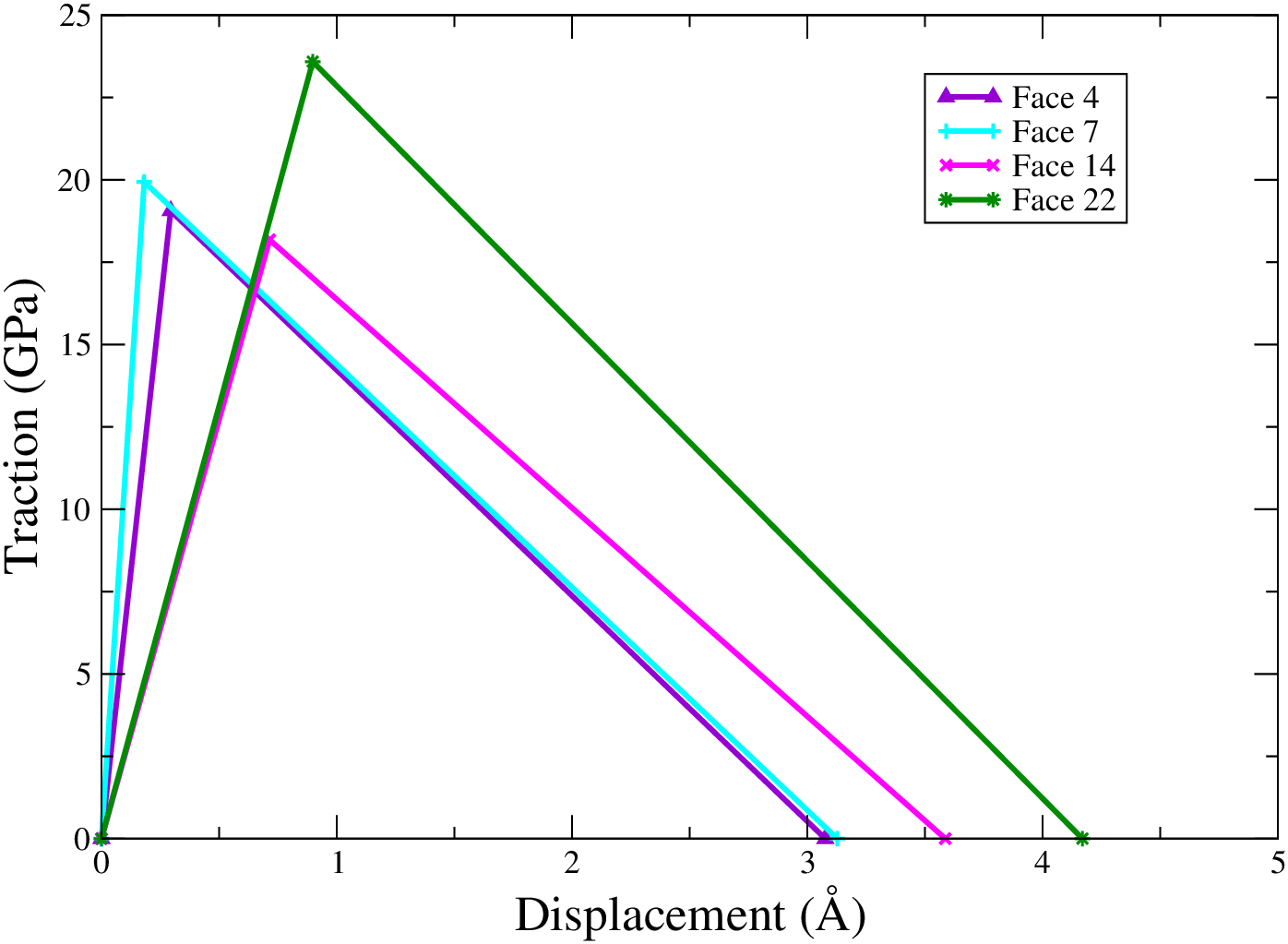}}
\mycaption[Piecewise Linear Cohesive laws: Ordinary SW Silicon.]
{The same as Figure~\ref{fig:BilinearCohesiveBrittle} but for the original,
  ductile SW potential for silicon.}
\label{fig:BilinearCohesiveOrdinary}
\end{center}
\end{figure}

Because we are measuring the displacements 30 $\mbox{\AA}$ from the actual
boundary, we need to subtract off the elastic response of the grain.  Because we
are using rollered boundary conditions, there is no Poisson-effect contraction,
and the relevant component of the elastic tensor is $C_{1111}$, describing the
strain normal to the grain boundary.  The elastic tensor for the rotated crystal
is found by rotating the elastic constants found in
section~\ref{sec:measuring_elastic_constants} by the same rotation matrix that
describes the rotation of the lattice vectors in each grain
\begin{equation}
C'_{1111} = R_{1i}R_{1j}R_{1k}R_{1l}C_{ijkl}.
\end{equation}
We must then combine $C'_{1111}$ from each grain such that the stress in each
grain is equal (analogous to springs in series)
\begin{equation}
\sigma = C^{(1)}_{1111} \frac{d_1}{W} = C^{(2)}_{1111}
\frac{d_2}{W} = C^{\mathrm{eff}}_{1111} \frac{d_1+d_2}{2W}
\end{equation}
\begin{equation}
C^{\mathrm{eff}}_{1111} = 2/(\frac{1}{ C^{(1)}_{1111}} + \frac{1}{ C^{(2)}_{1111}})
\end{equation}
where $d_1$ and $d_2$ refer to the displacement in each grain and $W$ is the
width of each grain.  The displacement near the grain boundary is then given by
\begin{equation}
d_{gb} = 2 W \epsilon - \frac{\sigma}{C^{\mathrm{eff}}_{1111}}  2 (W - W_{gb})
\label{eqn:subtract_elastic}
\end{equation}
where $W_{gb}$ represents a finite width associated with the interface and
$\epsilon$ is the external, normal strain.  Since,
in our system, the grain boundary is more stiff than the perfect crystal for the
silicon geometries we have studied, this finite width is necessary so
that~\eref{eqn:subtract_elastic} does not give a negative value.
Figure~\ref{fig:CohesiveBrittleCorrected} shows the result of applying this
correction to the data show in Figure~\ref{fig:CohesiveBrittleRaw}. The initial
stiffness is then given by the peak stress divided by the displacement at peak
stress.  The final displacement is set such that the Griffith criterion is met,
i.e. such that the area under the curve is equal to the difference between the
final surface energies of the broken grain and the initial energy of the grain
boundary interface, $\lambda_c = 2(\gamma - \gamma_{gb})/\sigma_c$.
Figures~\ref{fig:BilinearCohesiveBrittle} and \ref{fig:BilinearCohesiveOrdinary}
show the final, piecewise linear cohesive laws that are used by the CZM
simulations. 

For the perfect crystal, we can simply scale the cohesive law to a width equal
to the finite width used to process the grain boundary cohesive laws since we do
not need to separate the behaviour of the bulk from the behaviour of an
interface.  This has the effect of preserving the non-linear elastic response.
The non-linear elastic response of the bulk is not separated from the response
of the interface for the case of grain boundaries, since the elastic response of
the bulk that we subtract off is assumed to be linear. Since the grain
boundaries have a lower fracture stress than the perfect crystal, nonlinear
effects in the bulk are less important.

In principle, two boundaries for which the grains have been swapped (such as
faces 0\&1, 2\&6, 3\&5, 20\&21 as shown in Figure~\ref{fig:CubeInCubeFaces})
should have the same overall structure and therefore have the same cohesive law.
In practice, when simulating a finite region of a grain boundary,
\textit{microparameters} (the choice of section of the interface, the translations of the
grains relative to one another, and the cutting plane of each lattice
orientation) alter the grain boundaries that have the same
\textit{macroparamters} (grain orientation) or would otherwise be the same by
symmetry. The differences between the cohesive laws for the pairs 0\&1, 2\&6,
3\&5, 20\&21 in figures~\ref{fig:BilinearCohesiveBrittle}
and~\ref{fig:BilinearCohesiveOrdinary} indicate the scope of this effect, of
order 10\% (much smaller than the discrepancy between atomistic and continuum
simulations, which we will observe in section~\ref{sec:comparison}).

\section{Fully Atomistic Simulation}
\label{sec:fully_atomistic_cube_in_cube}
The fully atomistic model is run with a software package called Overlapping
Finite Elements and Molecular Dynamics (OFEMD) which is described in detail 
in\inThesis{ chapter~\ref{chapter:ofemd}}\inPaper{~\onlinecite{coffmanthesis,
URL:OFEMD}}.  OFEMD uses the
\textit{DigitalMaterial}~\cite{DigitalMaterial} library 
to run atomistic simulations of any geometry within a finite element mesh.
OFEMD uses the mesh information to fill each material region with atoms in the
given lattice orientation and set up contrained zones as described in
section~\ref{sec:MeasuringCohesiveLawForCubeInCube} to simulate rollered
boundary conditions.  The fully atomistic
simulation uses the same kinematic boundary conditions as the FE simulation:
a normal loading imposed on the upper face.  We
manually update the positions of the atoms in the constrained zones that are
adjacent to the upper face to impose this boundary condition, incrementing the
strain up to 15\% in 0.5\% strain increments, relaxing the atoms at each step.

\section{Cohesive Zone Model Comparison}
\label{sec:comparison}

We compare the fracture behaviour of atomistic and continuum FE simulations for
both the standard SW potential (which is ductile for single
crystal, intragranular fracture) and the modified, brittle SW
potential.  We use both potentials to check if discrepancies between atomistic
and continuum simulations could be due to ductility. We explore simulations of
two sizes (inner cube sizes of~10 and~20\AA) to check if discrepancies get
smaller in the continuum limit of larger specimens.  The interfacial cohesive
laws in each case were computed as in Figures~\ref{fig:BilinearCohesiveBrittle}
and~\ref{fig:BilinearCohesiveOrdinary}, from atomistic simulations with 30\AA\ grains.

Figures~\ref{fig:brittle10AngstromComparison}
through~\ref{fig:original20AngstromComparison} show the results of both the
atomistic and continuum simulations for both versions of SW
silicon and both length scales.  The colour scales denote
$\sigma_{zz}$, the vertical component of stress. The stresses for the atomistic
simulations were calculated using~\eref{eqn:atomicstress}.  The first row
of each figure shows the $xy$ center plane of the atomic simulation (roughly the
plane of fracture).  The second row shows the same plane of fracture for the CZM
simulation.  The third row shows the $xz$ center plane of the atomistic
simulation, illustrating the stresses around the fracture zone and the crack
opening.  The fourth row shows the $xz$ center plane of the CZM simulation. The
stress-free state, indicated by the colour blue, is an indication that decohesion
has occurred across the interface within that region.

We shall see that the atomistic simulations and the FE simulations differ in several
important respects. First, the FE simulations fracture overall at a higher
strain level. This might be a nucleation effect; the irregular atomic structures
at the external faces and internal edges and corners could be acting as
nucleation points for fracture in ways that are not reflected in the continuum
simulation. Second, the pattern of fracture---which interfaces break in which
order---is in some cases different for the two simulations. Some of these
differences are accidental; the system has inversion symmetries across the $xz$
and $yz$ planes that are broken only by the microparameter choices in the
grain-boundary cohesive law atomistic simulations and the fully atomistic cube-in-cube
simulations.  The FE simulation reflects the choice of microparameters chosen
in the cohesive law simulations while the fully atomistic simulation reflects another
choice of microparameters. Hence an atomistic simulation that breaks first along the
`front' edge is equivalent to a FE simulation breaking along the
`back'. Indeed, were we to use fully converged, infinite-system cohesive laws
such as the periodic boundary conditions used
in\inThesis{chapter~\ref{chapter:grain_boundaries}}
\inPaper{~\onlinecite{grain-boundary-geo}}, an ideal FE simulation would break
symmetrically.  However, this effect alone cannot account for the differences
between the FE simulations and the fully atomistic simulations.  We will see
that these differences are larger than the differences due to microparameter
choices (as observed in
figures~\ref{fig:BilinearCohesiveBrittle} and~\ref{fig:BilinearCohesiveOrdinary}).

\subsection{Brittle SW with a 10~\mbox{\AA} Inner Cube }

\begin{figure}
\begin{center}

\subfigure[11\%]{
\includegraphics[width=2.5cm]{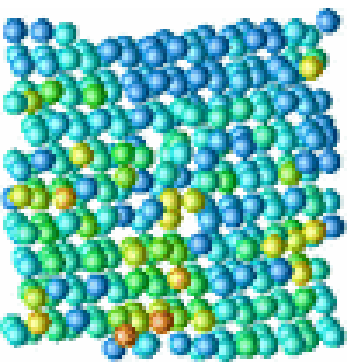}
\label{fig:brittle10atoms11xy}}
\subfigure[12\%]{
\includegraphics[width=2.5cm]{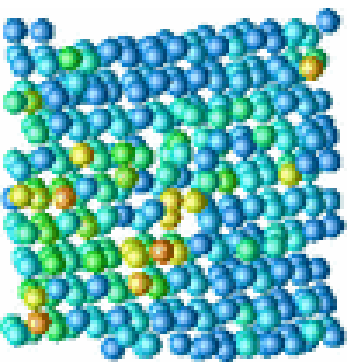}}
\subfigure[13\%]{
\includegraphics[width=2.5cm]{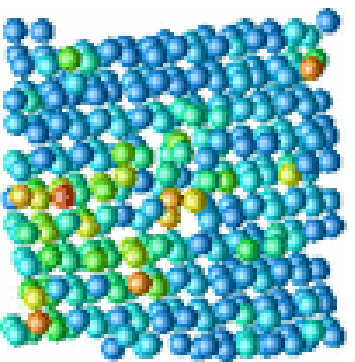}}
\subfigure[14\%]{
\includegraphics[width=2.5cm]{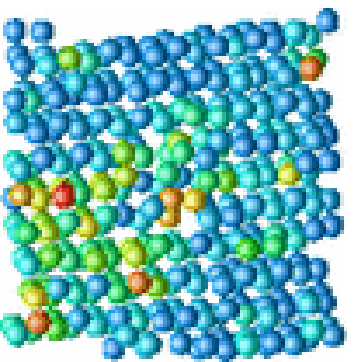}}
\subfigure[15\%]{
\includegraphics[width=2.5cm]{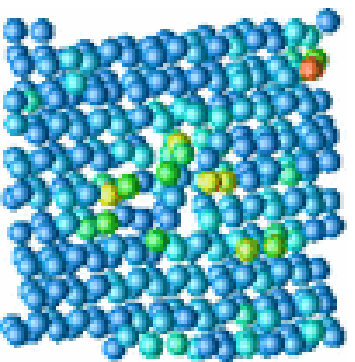}}

\subfigure[11.1\%]{
\includegraphics[width=2.5cm]{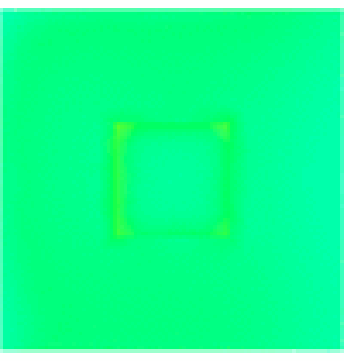}}
\subfigure[12.1\%]{
\includegraphics[width=2.5cm]{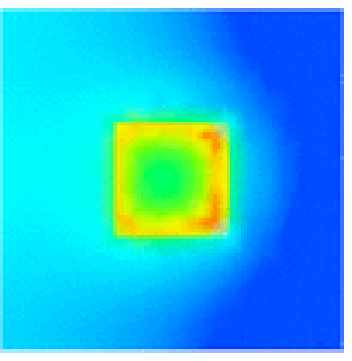}}
\subfigure[13.1\%]{
\includegraphics[width=2.5cm]{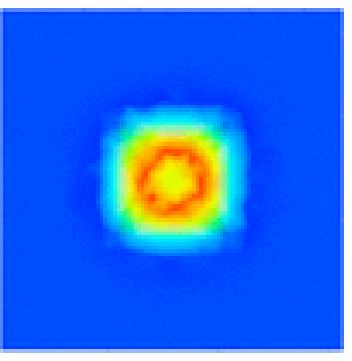}}
\subfigure[14.1\%]{
\includegraphics[width=2.5cm]{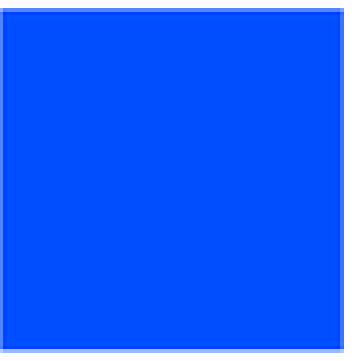}}
\subfigure[15.1\%]{
\includegraphics[width=2.5cm]{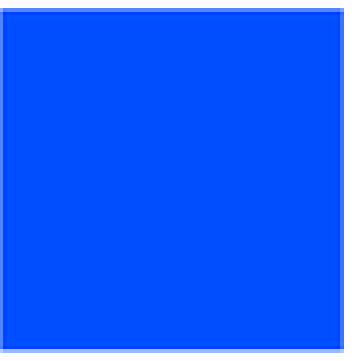}}

\subfigure[11\%]{
\includegraphics[width=2.5cm]{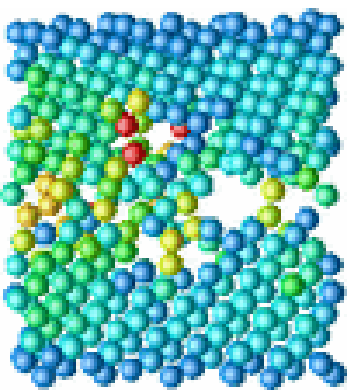}}
\subfigure[12\%]{
\includegraphics[width=2.5cm]{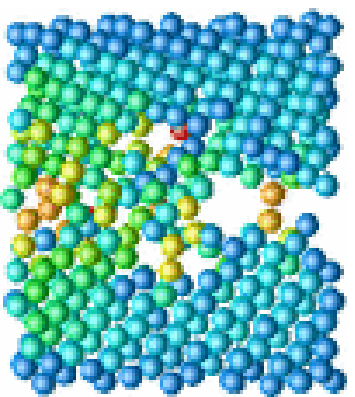}}
\subfigure[13\%]{
\includegraphics[width=2.5cm]{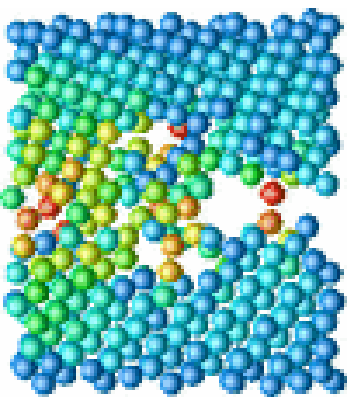}}
\subfigure[14\%]{
\includegraphics[width=2.5cm]{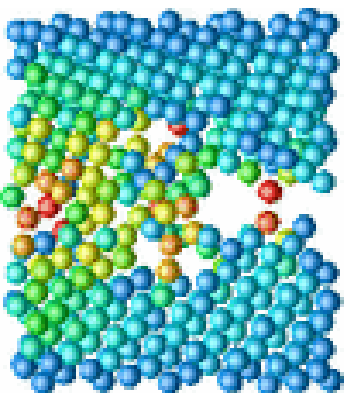}}
\subfigure[15\%]{
\includegraphics[width=2.5cm]{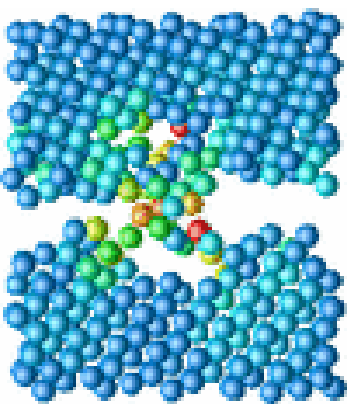}}

\subfigure[11.1\%]{
\includegraphics[width=2.5cm]{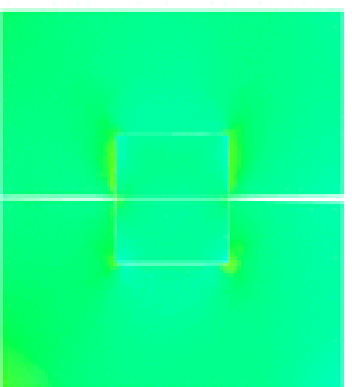}}
\subfigure[12.1\%]{
\includegraphics[width=2.5cm]{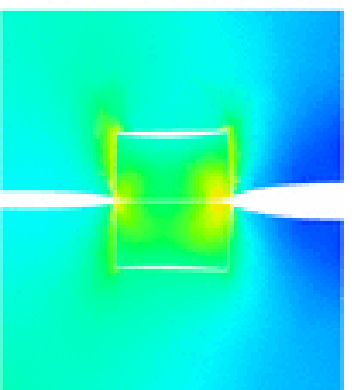}}
\subfigure[13.1\%]{
\includegraphics[width=2.5cm]{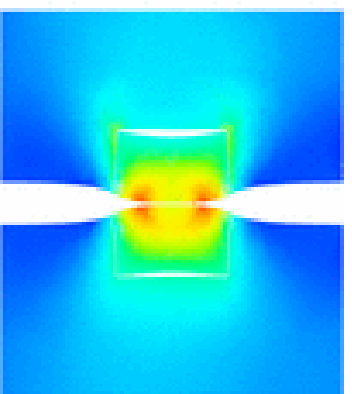}}
\subfigure[14.1\%]{
\includegraphics[width=2.5cm]{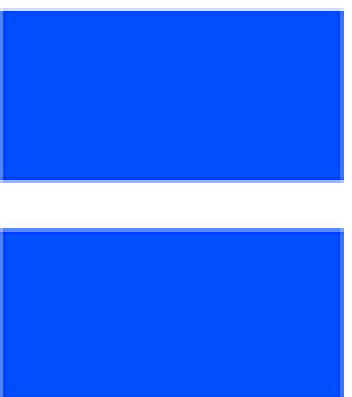}}
\subfigure[15.1\%]{
\includegraphics[width=2.5cm]{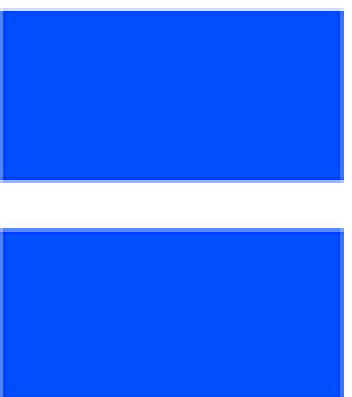}}

\includegraphics[width=6cm]{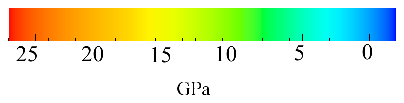}
\pagebreak[10]
\end{center}
\mycaption[Comparison of the Atomistic and CZM Simulations of the Cube-In-Cube
  with a 10 \mbox{\AA} Inner Cube, using Brittle SW
  Silicon.]{The top two rows are $\sigma_{zz}$ on the $xy$ center (fracture) plane.  The bottom
  two rows are $\sigma_{zz}$ on the $xz$ center, cross-sectional plane.}
\label{fig:brittle10AngstromComparison}
\end{figure}

Figure~\ref{fig:brittle10AngstromComparison} shows the comparison between
the smaller simulations of the brittle potential (an inner cube length of 10\AA,
with the brittle modification of the SW potential).
The atomistic simulation appears to begin to fracture at 11\% strain in the upper
right corner of the $xy$ plane in Figure~\ref{fig:brittle10atoms11xy} with the fracture spreading
across the right side and finally across the center plane, extracting, rather
than splitting, the inner
cube at 15\% strain.  At this small scale, the inner cube is amorphized during
the first relaxation step.

The finite element simulation begins to fracture on the right side as well between
11.1\% strain and 12.1\% strain, approximately where the atomistic simulation fractures.
The only feature which breaks the 90 degree
rotation symmetry for the finite element simulations are the differences in
cohesive laws. The finite element simulation
fractures slightly more rapidly, also ending by breaking through the inner cube
but at 14.1\% strain rather than 15\%.

\subsection{Brittle SW with a 20~\mbox{\AA} Inner Cube }

\begin{figure}
\begin{center}
\subfigure[8\%]{
\includegraphics[width=2.5cm]{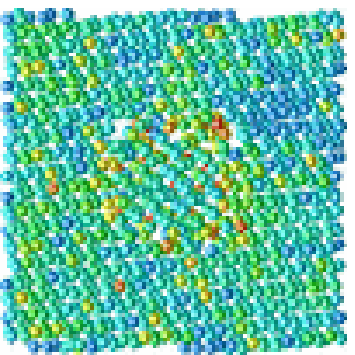}
\label{fig:brittle20atoms8xy}}
\subfigure[9\%]{
\includegraphics[width=2.5cm]{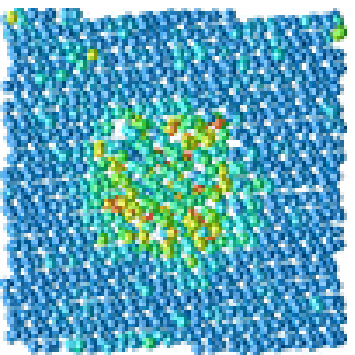}}
\subfigure[10\%]{
\includegraphics[width=2.5cm]{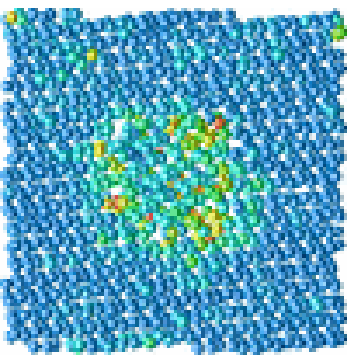}}
\subfigure[11\%]{
\includegraphics[width=2.5cm]{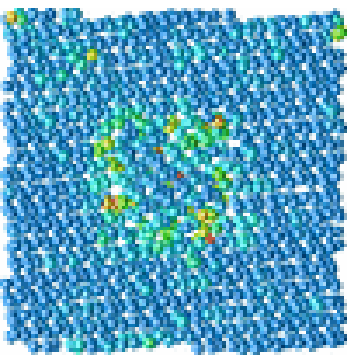}}

\subfigure[11.1\%]{
\includegraphics[width=2.5cm]{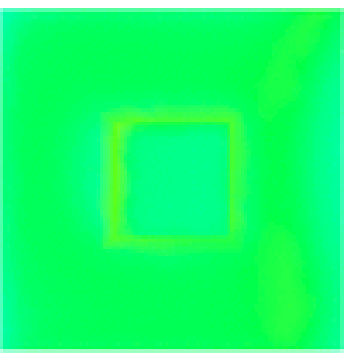}}
\subfigure[12.1\%]{
\includegraphics[width=2.5cm]{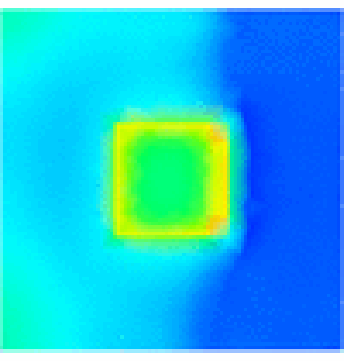}}
\subfigure[13.1\%]{
\includegraphics[width=2.5cm]{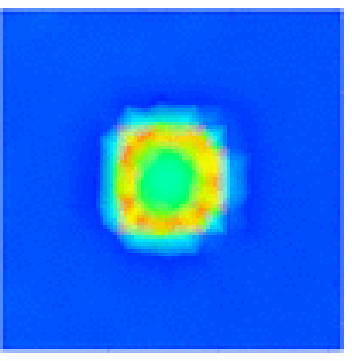}}
\subfigure[14.1\%]{
\includegraphics[width=2.5cm]{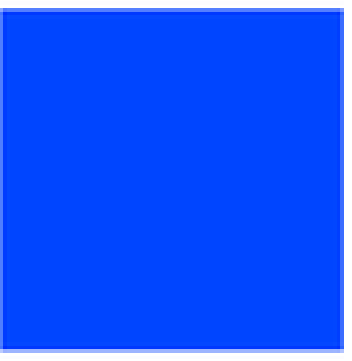}}

\subfigure[8\%]{
\includegraphics[width=2.5cm]{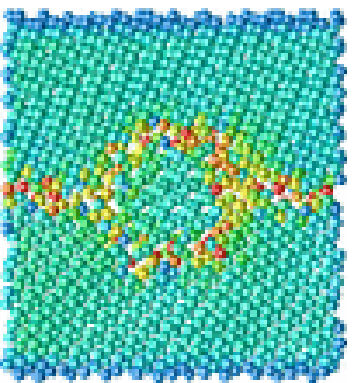}}
\subfigure[9\%]{
\includegraphics[width=2.5cm]{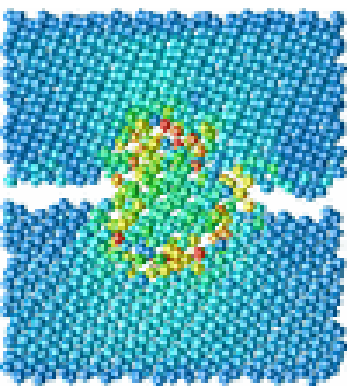}}
\subfigure[10\%]{
\includegraphics[width=2.5cm]{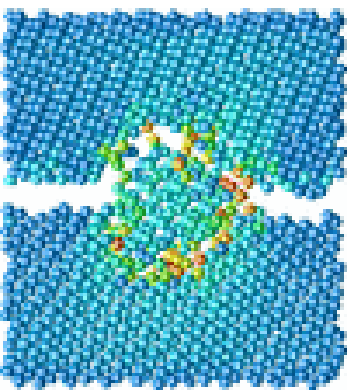}
\label{fig:brittle20atoms10xz}}
\subfigure[11\%]{
\includegraphics[width=2.5cm]{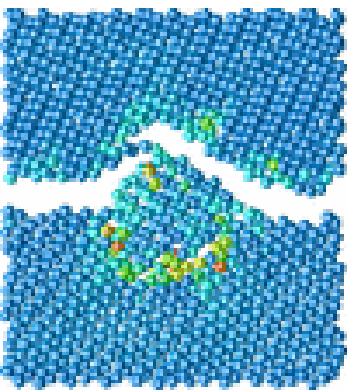}}

\subfigure[11.1\%]{
\includegraphics[width=2.5cm]{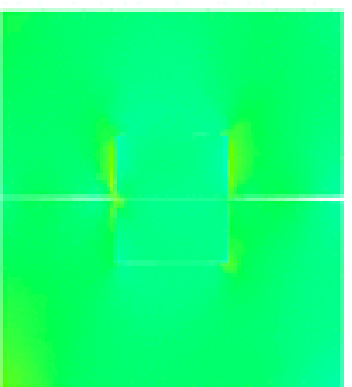}}
\subfigure[12.1\%]{
\includegraphics[width=2.5cm]{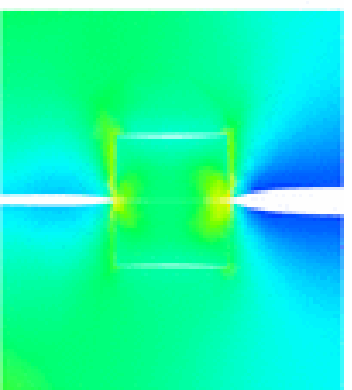}}
\subfigure[13.1\%]{
\includegraphics[width=2.5cm]{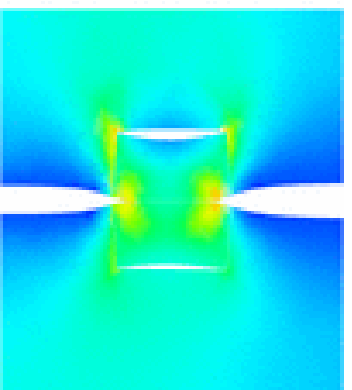}
\label{fig:brittle20fem13xz}}
\subfigure[14.1\%]{
\includegraphics[width=2.5cm]{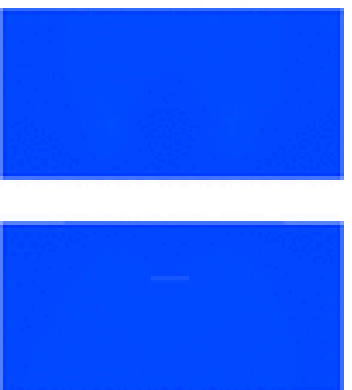}}

\includegraphics[width=6cm]{cube_in_cube_figures/10AngstromBrittleBeta1/colorscalestress}
\end{center}
\mycaption[Comparison of the Atomistic and CZM Simulations of the Cube-In-Cube
  with a 20 \mbox{\AA} Inner Cube, using Brittle SW
  Silicon.]{ The top two rows are $\sigma_{zz}$ on the $xy$ center (fracture) plane.  The bottom
  two rows areare $\sigma_{zz}$ on the $xz$ center, cross-sectional plane.}
\label{fig:brittle20AngstromComparison}
\end{figure}

For the 20 \mbox{\AA} length scale atomistic simulations
(Figure~\ref{fig:brittle20AngstromComparison}), fracture also begins at the upper
right corner of the $xy$ plane in Figure~\ref{fig:brittle20atoms8xy}, however
fracture begins noticeably earlier at 8\% strain and propagates through the
center plane more rapidly.  At 9\% strain, the atomistic simulation is
comparable to the continuum simulation at 13\% strain, with the center plane,
excluding the inner cube, cracked through.  This more rapid fracture of the
atomistic simulation could be due to microstructure differences, but could also
be due to the larger system size. A larger system height means there is more
energy stored in elastic strain per unit area of interface.  Once a given region
reaches the maximum stress that it can sustain, it snaps open.  With a smaller
system size, the opening of the interface is controlled since the constrained
zones are closer.
This is related to the effect described in
\inThesis{section~\ref{sec:Width}}\inPaper{\onlinecite{grain-boundary-geo}}
where larger systems
effectively approach fixed force boundary 
conditions\inThesis{(figures~\ref{fig:WidthTest-CohesiveLaw} 
and~\ref{fig:WidthTest-Atoms})}.

Both the atomistic simulation and the finite element simulation begin to decohere at
the upper face of the inner cube (compare Figure~\ref{fig:brittle20atoms10xz}
with the slight blue decohered region above the inner cube in
Figure~\ref{fig:brittle20fem13xz}). However, the FE simulation ultimately
decoheres at the center plane instead. In the atomistic simulation, we also see
a competition between cracking at the top of the inner cube and cracking through
the center plane.  Ultimately, the crack propagates partially through the inner
cube at an angle, reaching the top of the inner cube.  This effect cannot be
replicated in the finite element simulation because it did not have interface
elements in position to crack at this angle.

\subsection{Original SW with a 10~\mbox{\AA} Inner Cube }

\begin{figure}
\begin{center}
\subfigure[10\%]{
\includegraphics[width=2.5cm]{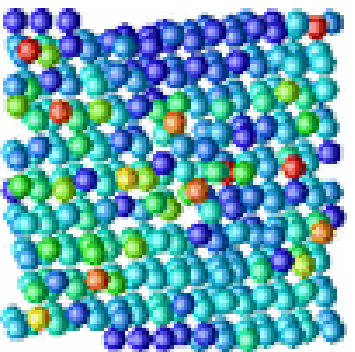}}
\subfigure[11\%]{
\includegraphics[width=2.5cm]{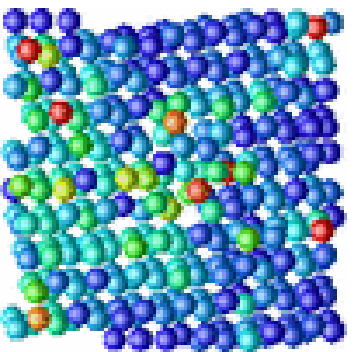}}
\subfigure[12\%]{
\includegraphics[width=2.5cm]{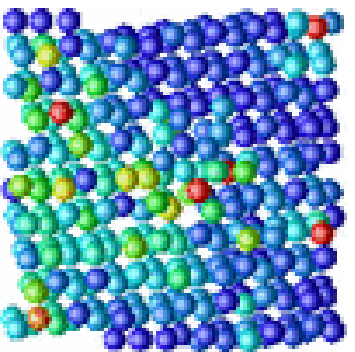}}
\subfigure[13\%]{
\includegraphics[width=2.5cm]{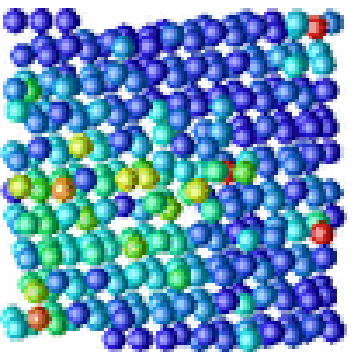}}
\subfigure[14\%]{
\includegraphics[width=2.5cm]{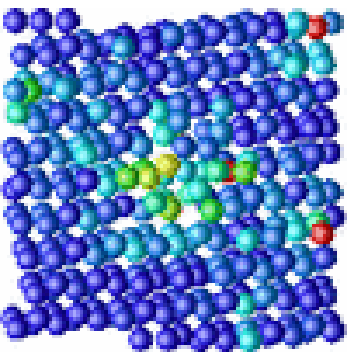}}

\subfigure[24.1\%]{
\includegraphics[width=2.5cm]{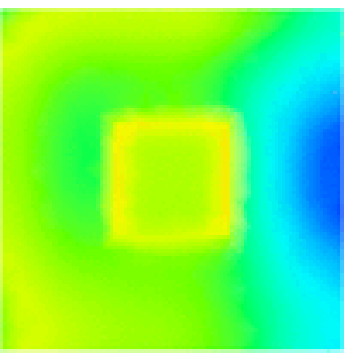}}
\subfigure[26.1\%]{
\includegraphics[width=2.5cm]{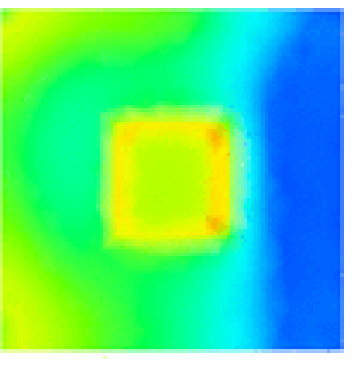}}
\subfigure[28.1\%]{
\includegraphics[width=2.5cm]{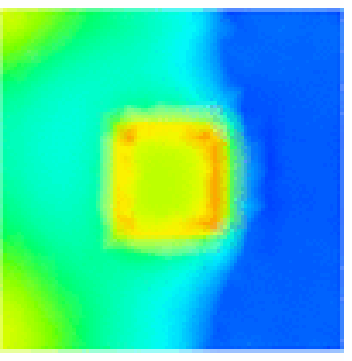}}
\subfigure[30\%]{
\includegraphics[width=2.5cm]{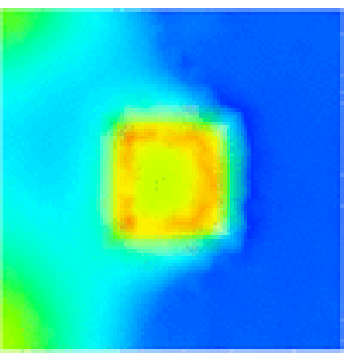}}

\subfigure[10\%]{
\includegraphics[width=2.5cm]{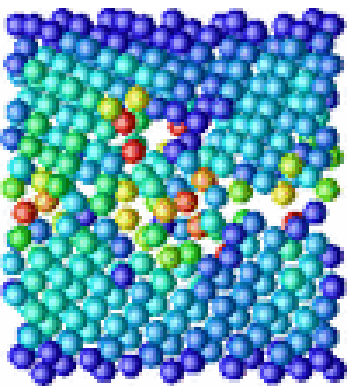}}
\subfigure[11\%]{
\includegraphics[width=2.5cm]{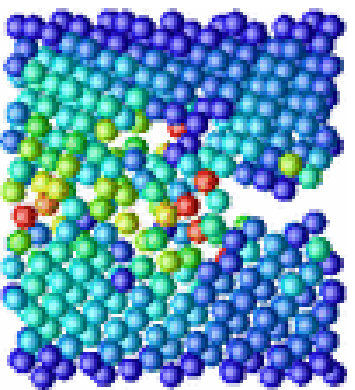}}
\subfigure[12\%]{
\includegraphics[width=2.5cm]{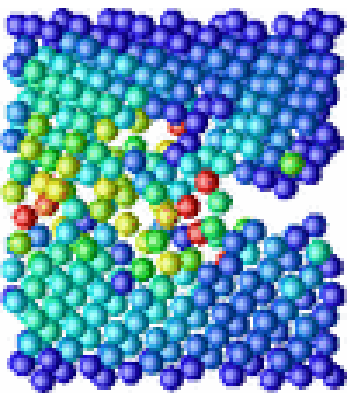}}
\subfigure[13\%]{
\includegraphics[width=2.5cm]{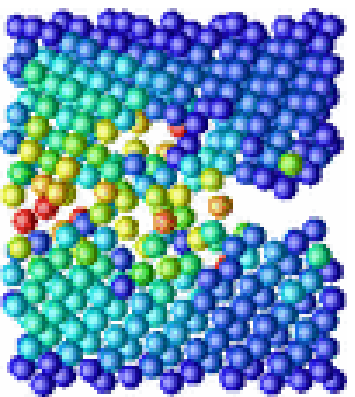}}
\subfigure[14\%]{
\includegraphics[width=2.5cm]{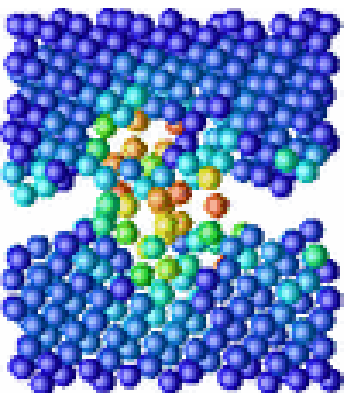}}

\subfigure[24.1\%]{
\includegraphics[width=2.5cm]{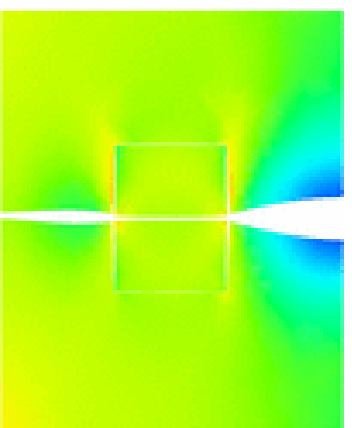}}
\subfigure[26.1\%]{
\includegraphics[width=2.5cm]{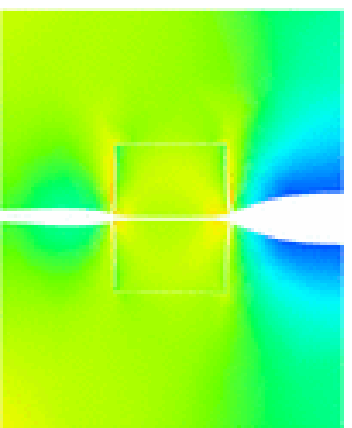}}
\subfigure[28.1\%]{
\includegraphics[width=2.5cm]{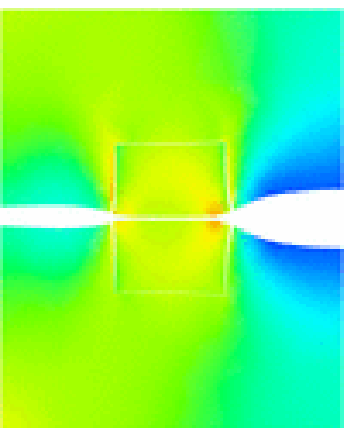}}
\subfigure[30\%]{
\includegraphics[width=2.5cm]{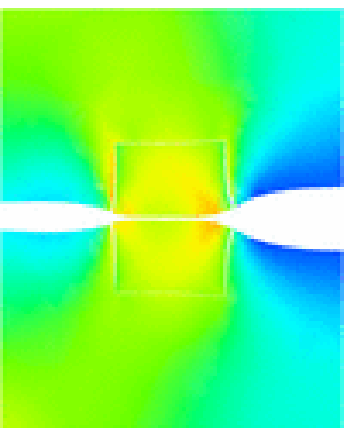}}

\includegraphics[width=6cm]{cube_in_cube_figures/10AngstromBrittleBeta1/colorscalestress}
\end{center}
\mycaption[Comparison of the Atomistic and CZM Simulations of the Cube-In-Cube
  with a 10 \mbox{\AA} Inner Cube, using Original SW
Silicon]{The top two rows are $\sigma_{zz}$ on the $xy$ center (fracture) plane.  The bottom two
rows are  $\sigma_{zz}$ on the $xz$ center, cross-sectional plane. }
\label{fig:original10AngstromComparison}
\end{figure}

For the original version of SW silicon (which is more ductile for
single-crystal fracture) with the smaller inner cube size (Figure~\ref{fig:original10AngstromComparison}), the atomistic simulations fracture at around 14-15\% strain,
similar to the fracture threshold seen for the brittle potential atomistic simulations
at that size. The 
continuum simulations, however, fracture at a much higher strain, 30\%
compared to 15\%, despite using cohesive-zone models derived from the original
potential.

The atomistic simulation begins to fracture at the top of the $xy$ plane
(10\% strain figure) and spreads along the right side (11, 12\% figures).  At
the end of the atomistic simulation (14\% strain), it has cracked through all but the
center cube. The CZM simulation
begins to fracture along the external edge along the side, and has also not cracked
through or around the inner cube at the conclusion of the simulation.

\subsection{Original SW with a 20~\mbox{\AA} Inner Cube }

\begin{figure}
\begin{center}
\subfigure[12\%]{
\includegraphics[width=2.5cm]{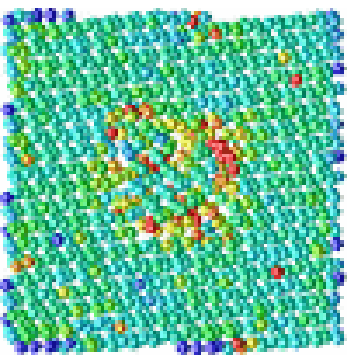}}
\subfigure[13\%]{
\includegraphics[width=2.5cm]{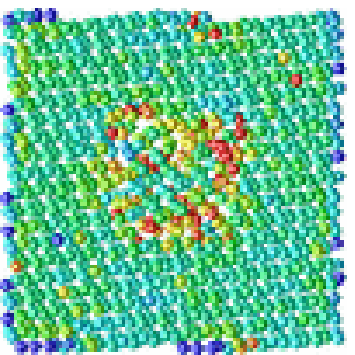}}
\subfigure[14\%]{
\includegraphics[width=2.5cm]{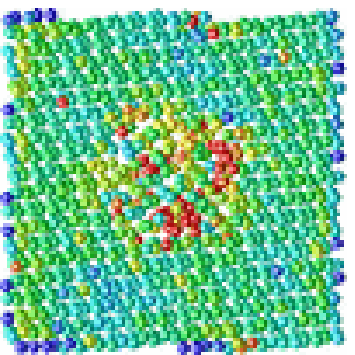}}
\subfigure[15\%]{
\includegraphics[width=2.5cm]{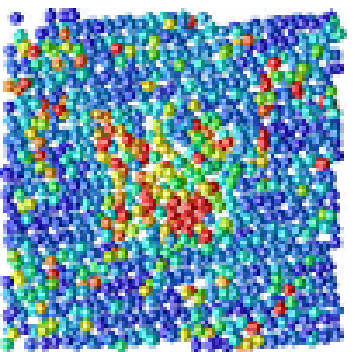}}

\subfigure[23.1\%]{
\includegraphics[width=2.5cm]{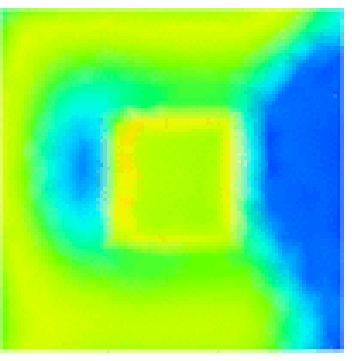}}
\subfigure[25.1\%]{
\includegraphics[width=2.5cm]{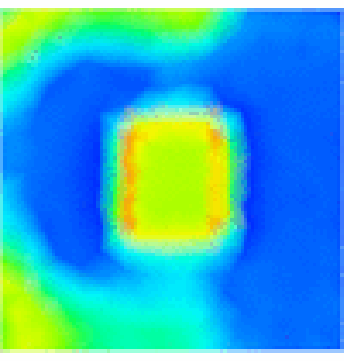}}
\subfigure[27.1\%]{
\includegraphics[width=2.5cm]{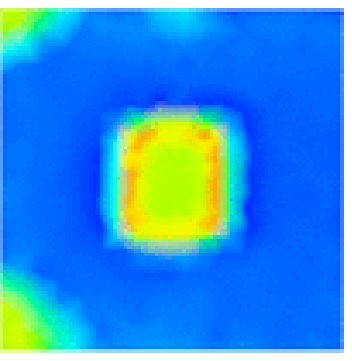}}
\subfigure[29.1\%]{
\includegraphics[width=2.5cm]{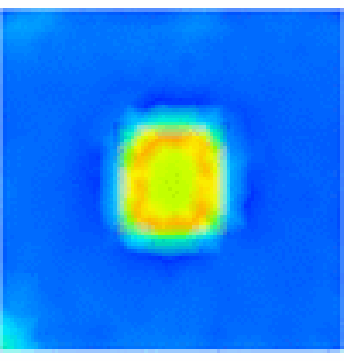}}

\subfigure[12\%]{
\includegraphics[width=2.5cm]{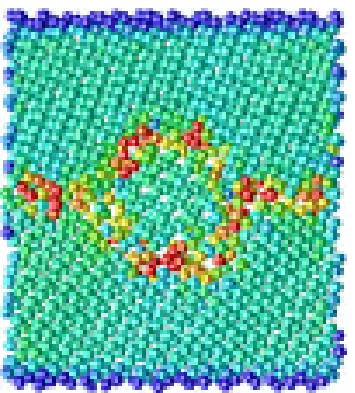}}
\subfigure[13\%]{
\includegraphics[width=2.5cm]{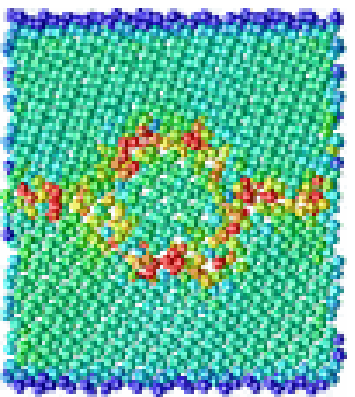}}
\subfigure[14\%]{
\includegraphics[width=2.5cm]{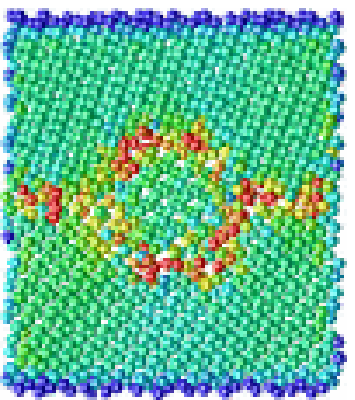}}
\subfigure[15\%]{
\includegraphics[width=2.5cm]{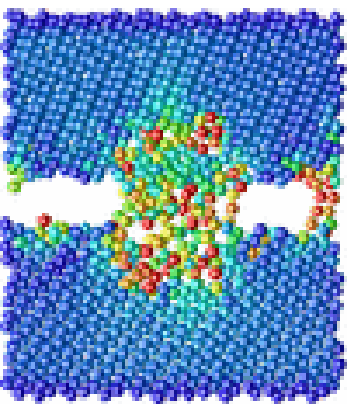}}

\subfigure[23.1\%]{
\includegraphics[width=2.5cm]{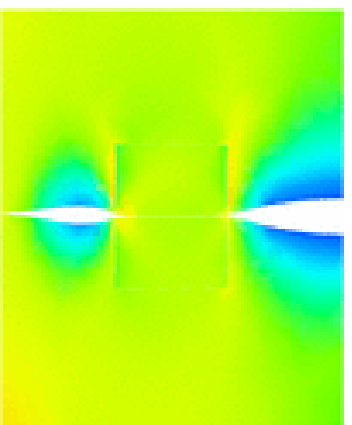}}
\subfigure[25.1\%]{
\includegraphics[width=2.5cm]{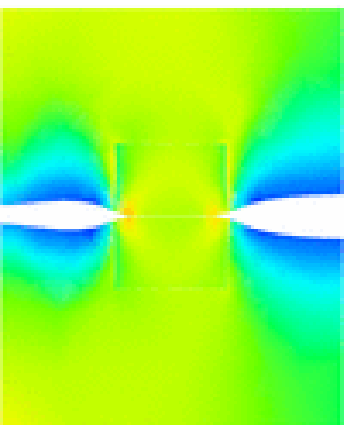}}
\subfigure[27.1\%]{
\includegraphics[width=2.5cm]{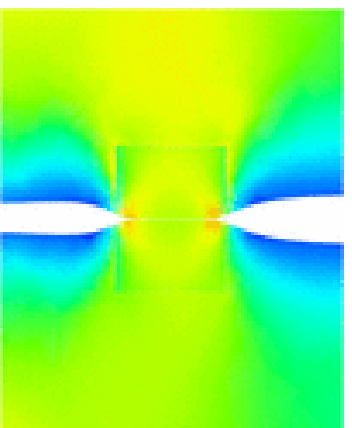}}
\subfigure[29.1\%]{
\includegraphics[width=2.5cm]{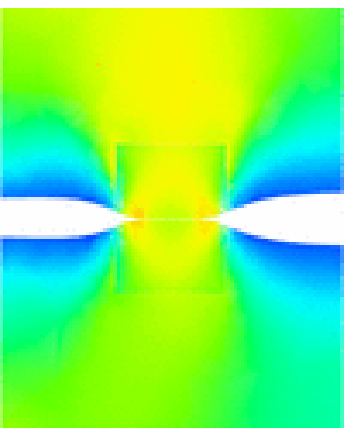}}

\includegraphics[width=6cm]{cube_in_cube_figures/10AngstromBrittleBeta1/colorscalestress}
\end{center}
\mycaption[Comparison of the Atomistic and CZM Simulation of the Cube-In-Cube
  with a 20 \mbox{\AA} Inner Cube, using Original SW
  Silicon.]{The top two rows are $\sigma_{zz}$ on the $xy$ center (fracture) plane.  The bottom
  two rows are $\sigma_{zz}$ on the $xz$ center, cross-sectional plane. } 
\label{fig:original20AngstromComparison}
\end{figure}

For the 20 \mbox{\AA} case (Figure~\ref{fig:original20AngstromComparison}), the ductile atomistic simulation again fractures at a much
lower stress than does the CZM simulation. The atomistic simulation
fractures through all but
the center cube very rapidly between 14\% and 15\% strain, reflecting again the
effective soft-spring fixed-stress fracture conditions from the larger system
size; the CZM simulation fractures more gradually, showing a sweep from right to left.
The behaviour of the CZM simulation is similar to that of the 10~\mbox{\AA} case with
fracture beginning on the right side and slowly propagating through the center
plane.


\section{Conclusion}

We have described a method for comparing finite element simulations of
polycrystal models to fully atomistic simulations of the same geometry. In the
finite element simulations, we used elastic constants and  cohesive laws for the
grain boundaries,
derived from the atomistic calculations.  We find fair
agreement between the two simulations in one case (the 10 \mbox{\AA} brittle
SW simulation) in terms of the strain at which the fracture begins
and the pattern of fracture. However, the more macroscopic, continuum, brittle
simulation showed poor agreement, where one would have naively expected
improved convergence.  The more ductile simulations showed poor agreement at
both length scales.

Some of the differences between the atomistic simulations and the finite element
simulations can be attributed to the difference in choice of microparameters
defining the grain boundary geometries (the location of the lattice origin with
respect to the interface).  Carefully matching these microparameters in future
comparisons can only partially correct for the differences -- there will
always be discreteness effects in atomistic simulations that cannot be
replicated in finite element simulations, due to the distortion of atoms at
interface corners and junctions of grain boundaries. These sites, which are not
described by cohesive laws, are often the locus for crack nucleation. In order
to extract the cohesive properties of complex local regions, we suggest the use
of direct, on-the-fly, locally atomistic simulations~\cite{coffmanthesis, URL:OFEMD}.
\inThesis{ In
Chapter~\ref{chapter:ofemd} we describe a method for decorating regions of a
finite element mesh with atoms to extract atomistic information about the
cohesive properties of geometrical features such as interface corners and grain
boundary junctions.}

\ack
This work was supported by NSF Grants No. ITR/ASP ACI0085969 and
No. DMR-0218475.  We also wish to thank Drew Dolgert, Paul
Stodghill, Surachute Limkumnerd, Chris Myers, and Paul Wawrzynek.

\section*{References}
\bibliographystyle{unsrt}
\bibliography{general}

\end{document}